\documentclass[twocolumn,showpacs,preprintnumbers,prb,aps,superscriptaddress, floatfix]{revtex4-2}
\usepackage{amsmath}
\usepackage{amssymb}
\usepackage[retain-unity-mantissa=false]{siunitx}
\usepackage{graphicx}
\usepackage{xcolor}
\usepackage[caption=false]{subfig}
\usepackage{placeins}
\usepackage[colorlinks=true, linkcolor=blue, citecolor=blue, filecolor=blue, urlcolor=blue,pdfproducer={.},pdfcreator={.}]{hyperref}
\usepackage{cleveref}
\usepackage{blindtext}
\usepackage{comment}
\usepackage{tikz}
\usepackage{verbatim}
\usepackage{ulem}
\usepackage{physics}
\usepackage[commandnameprefix=ifneeded, commentmarkup=uwave]{changes}
\usepackage{orcidlink}
\usepackage{lipsum}
\usepackage{placeins}
\usepackage{nicefrac}
\usepackage{mathtools}

\newcommand{\elel}{_\text{el-el}}
\newcommand{\elph}{_\text{el-ph}}
\newcommand{\ii}[1]{_\text{#1}}
\newcommand{\de}{{\rm d}E}

\makeatletter
\def\@fnsymbol#1{\ensuremath{\ifcase#1\or *\or \ddagger\or
   \mathsection\or \mathparagraph\or \|\or **\or \dagger\dagger
   \or \ddagger\ddagger \else\@ctrerr\fi}}
\makeatother

\definechangesauthor[name=Chris, color=orange]{C}
\definechangesauthor[name=Markus, color=cyan]{M}
\definechangesauthor[name=Tobi, color=purple]{T}
\definechangesauthor[name=Sebastian, color=magenta]{S}
\definechangesauthor[name=Sebastian, color=teal]{B}

\begin{document}

\title{Interplay of non-local transport and local scattering during electron thermalization and spatial equilibration in laser-excited metals}
\author{Markus Uehlein \orcidlink{0000-0002-3193-3749}}
\email{uehlein@rptu.de}
\author{Tobias Held \orcidlink{0009-0009-8925-1810}}
\author{Christopher Seibel \orcidlink{0000-0003-1513-1364}}
\author{Sebastian T. Weber \orcidlink{0000-0002-9090-2248}}
\author{Baerbel Rethfeld \orcidlink{0009-0008-9921-4127}}    
\email{rethfeld@rptu.de}
\affiliation{RPTU University Kaiserslautern-Landau, Department of Physics and State Research Center OPTIMAS, Kaiserslautern, Germany}

\begin{abstract}
    Ultrafast laser excitation of metals induces electronic nonequilibrium both in space and locally in the energy distribution.
    The subsequent dynamics are governed by the interplay between non-local transport and local scattering of hot electrons, yet combined microscopic descriptions of these processes remain sparse.
    Here, we disentangle the influence of these processes on thermalization using a reformulation of the Boltzmann transport equation in energy space that consistently describes both spatial equilibration and scattering through full collision integrals.
    Our results reveal that transport accelerates the apparent thermalization observed at the irradiated surface by removing athermal carriers, while the same spatial redistribution delays complete equilibration of the full electron system.
    We analyze the experimentally accessible energy-dependent dynamics at the front and back surface and find that the dominant process varies, depending on both position and on the energetic window.
    Overall, our work improves the understanding of the interplay of electronic nonequilibrium processes occurring in optically thick laser-driven systems with relevant implications for future electronic applications.
\end{abstract}

\date{\today}

\maketitle

\section{Introduction}

Ultrafast laser excitation provides a powerful tool for controlling and investigating nonequilibrium dynamics in solids.
Applications range from fundamental studies in plasmonics or solid state physics to technological fields, such as electronic and spintronic devices~\cite{Bauer2015,Caruso2026,Zutic2020,Rethfeld2017, Schirato2023}.
For thin, homogeneously excited samples, the laser excites hot electrons and drives the electron system far from energetic equilibrium. 
Subsequent scattering processes thermalize the electron system and transfer energy to the lattice.
For samples with a thickness larger than the optical penetration depth, those local scattering processes compete with non-local transport of energy and charge carriers.
The exploitation of the excited hot carriers for applications is critically limited by the timescale on which they thermalize~\cite{Brongersma2015, Bauer2015}.
Thus, understanding the different mechanisms after ultrafast laser-excitation has been a long-standing focus of both experimental~\cite{Sun1994, Hohlfeld1997, Liu2005, Bauer2015, Fann1992a, Obergfell2020, LloydHughes2021, Kuehne2022} and theoretical studies~\cite{Rethfeld2004b, Battiato2010, Seibel2026arxiv, Vanzan2024, Schirato2023, Caruso2026}.

Femtosecond-resolved techniques, such as time-resolved photoemission spectroscopy (PES), have been instrumental in tracking the evolution of nonequilibrium carrier populations~\cite{LloydHughes2021}.
In PES experiments, materials are excited with relatively low laser fluences, and the energy-resolved electron distributions are probed near the surface~\cite{Bauer2015, Fann1992a, LloydHughes2021, Kuehne2022}.
These populations evolve under the combined influence of intrinsic scattering and transport processes because the laser penetration depth in metals is typically only a few nanometers~\cite{palik, Dressel, Bevillon2018}.
As a result, the measured carrier dynamics are determined not only by local thermalization processes but also by the spatial redistribution of excited electrons across the depth of the sample~\cite{Bauer2015, Schmuttenmaer1994, Heckschen2023}.

A predictive theoretical description of these intertwined dynamics can, for example, be done with the Boltzmann transport equation, which is capable of describing the dynamics of the nonequilibrium distributions of electrons and phonons
~\cite{Rethfeld2002, Mueller2013PRB, Seibel2023, Caruso2022, Held2025collapse}, while 
in principle accounting for both, local scattering and spatial transport.

However, explicitly incorporating electron-electron scattering with realistic band structure effects is computationally demanding and is therefore typically performed in energy space~\cite{Knorren2000, Mueller2013PRB}.
In contrast, transport calculations require spatial resolution, which increases computational cost, and information about the direction of motion, which is not inherently included in the energy space representation~\cite{Knorren2000}.
Consequently, simultaneously treating both aspects remains highly challenging, and most theoretical studies rely on simplifying assumptions.
They either assume homogeneous heating~\cite{Seibel2023, Sun1994} and resolve scattering in detail, or retain transport while employing approximate scattering models, such as relaxation time approximations~\cite{Liu2005, Nenno2016, Hopkins2009b}.
Such approximations make it difficult to disentangle the contributions of intrinsic thermalization and transport on the measured carrier dynamics.

In this work, we present a spatially resolved Boltzmann framework that incorporates both local scattering and non-local transport. 
The approach combines full microscopic electron-electron and electron-phonon collision integrals with a two-branch description of electron motion normal to the surface.
This enables us to investigate the intertwined influence of scattering and transport on the evolution of nonequilibrium carrier populations, both at the excited surface and across the depth of the sample.
By directly combining scattering and electron motion, the entire range from ballistic to diffusive transport is implicitly covered.
We focus on describing the early stage of thermalization and spatial equilibration and thus restrict ourselves to the study of the evolving electron distribution.

Although the framework is suitable for treating arbitrary electronic densities of states, we apply it here to an aluminum-like free-electron metal in order to separate intrinsic effects from material-specific phenomena.
By comparing homogeneous with inhomogeneous Beer-Lambert excitation at the same absorbed energy, we quantify how transport processes modify the apparent thermalization dynamics near the surface.
We further analyze energy-resolved carrier densities in narrow spectral windows above the Fermi level, providing a direct connection to observables in PES experiments.
Thereby, we identify the influences of the various interaction mechanisms and analyze which mechanisms drive the dynamics in specific energy ranges.

\section{Methods}
When a thick metallic sample is irradiated by an ultrashort optical laser pulse, energy is absorbed by conduction electrons near the surface.
This process induces nonequilibrium both, in their local energy distribution, as well as in space due to the short laser penetration depths of few nanometers.
Subsequently, local scattering processes drive the distribution to a thermal state and transport processes drive spatial equilibration in the direction of light propagation ($z$-direction).

\begin{figure}
    \centering
    \includegraphics[width=.5\textwidth]{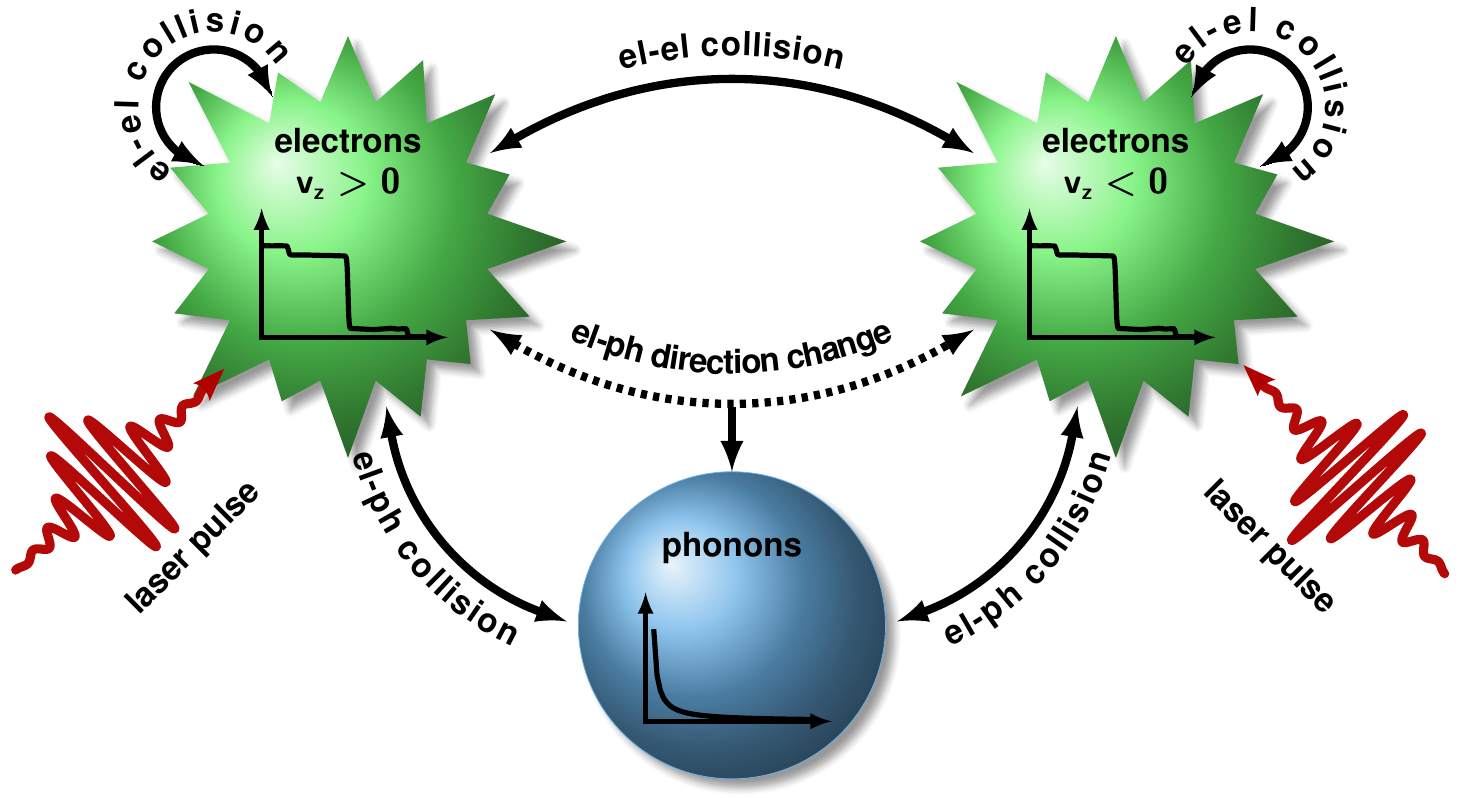}\\[.5cm]
    \includegraphics[width=.5\textwidth]{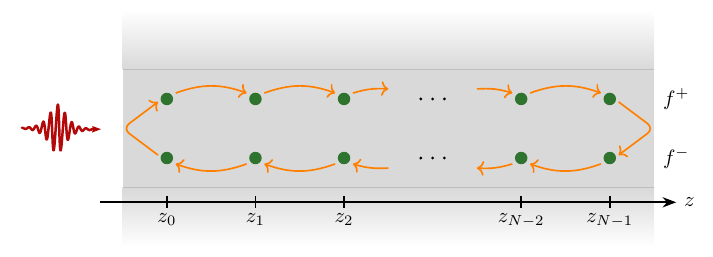}
    \caption{Schematic illustrations of the model structure.
    The upper scheme depicts the two electronic and one phononic subsystems and their couplings.
    The laser excites both electronic subsystems equally. 
    Solid arrows indicate an exchange of energy.
    The dashed arrow represents an exchange of electrons, which corresponds to a change in the direction of motion.
    The lower sketch shows the discretization of the sample and the different electron branches.
    The laser pulse excites the material from the surface of the solid at $z=0$.
    We trace at each point in spatial grid the time evolution of two electron distributions $f^\pm$ describing the electron population with velocities $v_z > 0$ or $v_z < 0$.
    Orange arrows indicate the direction of motion of the electrons.
    Due to reflections at the front and rear surfaces, the electrons change the direction of motion.
    For clarity, we omitted the arrows indicating changes in electron direction due to electron-phonon collisions.
    }\label{fig:panda}
\end{figure}

We characterize the electron system by its energy distribution $f_E \equiv f(E, t, z)$, which depends on energy $E$, time $t$, and depth in the material $z$.
The dynamics of the distribution is traced with the Boltzmann transport equation, which allows us to capture the athermal stage during and after ultrashort laser excitation.
We assume an isotropic momentum space, which allows for a transformation to energy space, and neglect external forces.
Then, the Boltzmann equation reads
\begin{align}
    \pdv{f_E}{t} + v \pdv{f_E}{z} &= \left.\pdv{f_E}{t}\right|_\text{laser}\!+ \left.\pdv{f_E}{t}\right|\elel + \left.\pdv{f_E}{t}\right|\elph \,,\label{eq:boltzmann}
\end{align}
with the velocity $v$ of the electrons.
Here, the left-hand side captures the spatiotemporal change of the distribution, while the collision terms on the right-hand side account for laser excitation and electron-electron and electron-phonon scattering, respectively.
In order to describe the equilibration of the energy gradient, it is necessary to be able to capture electron movement both into the bulk of the material and back to the excited surface.
This requires a tradeoff, since evaluating full electron-electron collision integrals with space-, energy- and momentum-resolution is not numerically feasible.
One approach is tracing the momentum-dependent electron distribution while approximating electron-electron scattering with a relaxation time description~\cite{Guenault1964, Kabanov2008, Nenno2016}.
However, the accuracy of this approximation is limited when it comes to energy-resolved dynamics~\cite{Uehlein2025, Roden2026}.
Therefore, in this work we restrict our description to the space- and energy-dependent distribution of electrons and use full microscopic Boltzmann collision integrals in \cref{eq:boltzmann}.
In order to resolve the different directions of motion, we project the momentum of the electrons onto the $z$-direction, divide the dispersion into two branches, and apply the \mbox{random-$k$} approximation~\cite{Penn1985} in each of these hemispheres.
One branch describes motion into the depth, $v_z>0$, and the other describes motion towards the excited surface, $v_z<0$.
We discretize both branches ('$\pm$') of the electron distribution~$f^\pm(E, t, z)$ on a spatial grid and evolve them in time according to \cref{eq:boltzmann}.
The energy-resolved density of states (DOS) of the material is equally distributed to both branches.
Our theoretical framework is schematically depicted in \cref{fig:panda}.
The upper part depicts both electron branches as separate systems, together with the phonon system and their respective local couplings, while the lower part  shows the non-local connection between the distributions of both branches $f^\pm$ at the points in space.

Next, we explain the applied formulations of the microscopic collision integrals on the right-hand side of \cref{eq:boltzmann}.

\subsection{Laser excitation\label{subsec:excitation}}

We are capturing the laser excitation similar to the description provided in Ref.~\citenum{Uehlein2025}, which offers an approach for calculating an athermal carrier distribution when a given energy density is absorbed.

The electronic distribution at a given energy~$E$ changes in two different ways, when a photon of energy~$\hbar\omega$ is absorbed.
The distribution decreases at energy~$E$, when an electron is excited to energy~$E + \hbar\omega$, generating a hole~('$h$') at energy~$E$, or increases at this energy, when an electron ('$e$') is excited from a lower energy state \mbox{$E - \hbar\omega$}.
Both processes add up to the change of the energy distribution
\begin{align}
    \left.\pdv{f_E}{t}\right|\ii{laser} &= \left.\pdv{f_E}{t}\right|_{h} + \left.\pdv{f_E}{t}\right|_{e}\,,
\end{align}
which enters \cref{eq:boltzmann}.
Here, as in \cref{eq:boltzmann}, we use the abbreviation $f_E=f^\pm(E, t, z)$.

Based on Fermi's golden rule, we can express the change of occupation at energy $E$ due to the loss and gain of electrons as
\begin{align}
    \left.\pdv{f_E}{t}\right|_{h\phantom{e}}\! &= - \frac{2\pi V}{\hbar}\,D_{E+\hbar\omega}\,|M_{h}|^2\,f_E (1-f_{E+\hbar\omega})\,,\\
    \left.\pdv{f_E}{t}\right|_{e\phantom{h}}\! &= \phantom{-} \frac{2\pi V}{\hbar}\,D_{E-\hbar\omega}\,|M_{e}|^2\,f_{E-\hbar\omega}(1-f_E)\,,
\end{align}
where the factors containing the energy distribution represent the probability of finding available states and account for Pauli blocking, respectively.
Here, $D_E \equiv D(E)$ denotes the electronic DOS and $V$ the volume of the unit cell.
We assume the two effective matrix elements $|M_{h}|^2$ and $|M_{e}|^2$ to be independent of energy.

We calculate the matrix elements at every point in time and space, taking into account the absorption of a given laser power density and considering particle conservation, meaning that an equal number of holes and electrons must be generated at all times (c.f. Ref.~\citenum{Uehlein2025}).
Therefore, we calculate the first moment of the distribution change, which represents the change of the electron density caused by the two processes.
Particle conservation then requires 
\begin{align}
- \int \left.\pdv{f_E}{t}\right|_{h}\,D_E\,\de &= \int \left.\pdv{f_E}{t}\right|_{e}\,D_E\,\de . \label{eq:primary_electrons_particle_cons}
\end{align}

The energy input at time $t$ in depth $z$ in each electron branch can be expressed as
\begin{align}
    s(t, z) &= \int  \left.\pdv{f_E(t, z)}{t}\right|\ii{laser}\,D_E\,E\,\de\,,\label{eq:laser_power_density}\\
    &\stackrel{!}{=} \frac{1}{2}\alpha\,I_0(t)\,\exp(-\alpha z)\,, \label{eq:laser_power_density_2}
\end{align}
where $s(t, z)$ is the laser power density. 
It is given by the laser intensity at the surface $I_0(t)$ and the absorption coefficient $\alpha$ and decays exponentially in the material, according to Beer-Lambert's law.
The factor of $\nicefrac{1}{2}$ distributes the laser power to both branches of the dispersion.

\subsection{Electron-electron scattering\label{subsec:el-el}}
We use a FCI within the random-$k$ approximation to describe electron-electron scattering~\cite{Roden2026}
\begin{align}
    \left.\pdv{f_{E}}{t}\right|_{\text{el-el}}&= \frac{4\pi}{\hbar} \, \iint D_{E'}D_{\varepsilon}D_{\varepsilon + \Delta E}\mathcal{F}_E\abs{M_{\text{ee}}}^2 \dd \varepsilon \dd E' \,, 
\end{align}
with the collision functional
\begin{align}
    \mathcal{F}_E &= f_{E'}f_{\varepsilon+\Delta E}(1-f_{E})(1-f_{\varepsilon})\nonumber\\
    &\phantom{=}~-f_{E}f_{\varepsilon}(1-f_{E'})(1-f_{\varepsilon+\Delta E}) \,,
\end{align}
capturing the probability of scattering accounting for occupied and unoccupied states, where $\Delta E = E - E'$ is the energy exchanged due to the collision with another electron.
Here, $\epsilon$ and $\epsilon+\Delta E$ are the energies of the initial and final state of the collision partner.
The matrix element $|M_{\text{ee}}|^2$ captures the quantum mechanical transition probability, which we derived using spherically averaging in a plane wave approach with a screened Coulomb potential~\cite{Roden2026}.
The resulting matrix element depends only on the screening parameter, which depends on the electron distribution and  is thus time-dependent~\cite{DelFatti2000,Mueller2013PRB}.

We also allow for electron-electron scattering between the directional branches, albeit without particle exchange, as in small-angle Coulomb scattering the direction of motion of an electron is mostly conserved.
Therefore, we evaluate two electron-electron collision integrals for each branch and spatial point.
In the case of the scattering of two electrons from different branches, $f_E$ and $f_{E'}$ belong to one branch, while $f_{\varepsilon}$ and $f_{\varepsilon+\Delta E}$ belong to the other branch.

\subsection{Electron-phonon scattering\label{subsec:el-ph}}
We use an FCI to capture electron-phonon scattering as described in Ref.~\citenum{Held2025a}:
\begin{align}
    \left.\pdv{f_{E}}{t}\right|_{\text{el-ph}}&= \frac{2 V \pi^3}{\hbar k_E} \int \sum_\pm \frac{D_{E\pm E_q}}{k_E+q} \frac{D^\text{ph}_{E_q}}{q} \mathcal{F}^\pm_E \abs{M_{\text{ep}}}^2 \Xi\ii{ep} \dd{E_q}\,,
\end{align}
where $V$ is the volume of the unit cell, $k_E$ is the electron wavenumber, and $D^\text{ph}_{E_q}$ is the phonon DOS at energy $E_q$, with the phonon wavenumber $q$.
The collision functional in this case is
\begin{align}
    \mathcal{F}^\pm_E &= f_{E\pm E_q} (1-f_{E}) \left(g + \frac{1}{2} \pm \frac{1}{2}\right)\nonumber\\
    &\phantom{= } - f_{E} (1-f_{E\pm E_q}) \left(g + \frac{1}{2} \mp \frac{1}{2}\right)\,,
\end{align}
where $g = g(E_q)$ is the phonon distribution.
For details regarding the transition matrix element $|M_{\text{ep}}|^2$ and the $\Xi\ii{ep}$-function that ensures momentum conservation, see Ref.~\citenum{Held2025a}. 

When an electron interacts with a phonon, the electron can either maintain or change its direction of motion.
In the latter case, $f_E$ and $f_{E\pm E_q}$ belong to different branches.
Thus, we also evaluate two electron-phonon FCI for each branch and point in space.

\subsection{Numerical details}\label{sec:details}
The calculations were performed with the \textit{monstr} simulation framework~\cite{Uehlein2026arxiv}.

As explained above, we only resolve the electron motion in $z$-direction.
Furthermore, we assume that all electrons travel in arbitrary directions at the same effective velocity, the Fermi velocity $v\ii{F}$.
We thus calculate the projection onto the $z$-axis of the velocity by spherically averaging over a hemisphere, see appendix~\ref{sec:vel}, and obtain $v_z^+ = 0.5\,v\ii{F}$ and $v_z^- = -0.5\,v\ii{F}$ for the different branches.
We divide our sample of length $L=N\Delta z$ into $N$ grid cells of width $\Delta z$ and place a discretization point at the center of each cell, such that $z_i = i_z \Delta z + \Delta z/2$ with $i_z \in \{0, N-1\}$.

We use a finite difference scheme to calculate the spatial derivative of the distribution in \cref{eq:boltzmann}.
We choose backward difference, 
\begin{align}
    \pdv{f^+(z)}{z} = \frac{f^+(z) - f^+(z-\Delta z)}{\Delta z}\,,
\end{align}
for calculating the transport into the depths of the material, while using a forward difference,
\begin{align}
    \pdv{f^-(z)}{z} = \frac{f^-(z+\Delta z) - f^-(z)}{\Delta z}\,,
\end{align}
for the transport back to the surface.
To capture the reflection at the surface, we replace the point outside the discretization with the corresponding point in the other branch of the distribution discretization, i.e.
\begin{align}
    \pdv{f^+(z_0)}{z} &= \frac{f^+(z_0) - f^-(z_0)}{\Delta z}\,,\\
    \pdv{f^-(z_{N-1})}{z} &= \frac{f^+(z_{N-1}) - f^-(z_{N-1})}{\Delta z}\,.
\end{align}
This combination of discretization, boundary conditions, and description of the gradient ensures energy and particle conservation.

We assume that 30\,\% of the electron-phonon collisions change the direction of the electrons. 
We confirmed that the qualitative results of this study remain robust against variations in this parameter; a brief discussion of its influence can be found in appendix~\ref{sec:neq}.
Furthermore, we verified that if an initial temperature gradient is given, our model predicts a temperature equilibration time throughout the sample that is comparable to that obtained with the well-known two-temperature model, see appendix~\ref{sec:ttm}.

\section{Results}

\subsection{Sample setup}
With our model, we study the influence of electronic transport on thermalization in a laser-excited metal with a free electron density of states $D(E) \propto \sqrt{E}$.
The parameters of the metal are chosen to match the properties of aluminum, which is well approximated by a free electron gas, as confirmed in appendix~\ref{sec:dos}.
We apply the average Fermi velocity $v\ii{F} = \SI{1.6}{\nano\meter\per\femto\second}$ of aluminum obtained from first principle calculations~\cite{Gall2016} and assume a constant absorption coefficient $\alpha = \SI{1.31e8}{\per\meter}$ for aluminum at \SI{800}{\nano\meter} excitation wavelength~\cite{Rakic1995}.
The phonons of the material are described with a triply degenerate Debye dispersion with an effective speed of sound of \SI{4167}{\meter\per\second}, which is the average of the longitudinal and the two transversal modes~\cite{CRC2005}.

We simulate the excitation and subsequent dynamics of a \SI{50}{\nano\meter} thick sample, which is discretized in \SI{1}{\nano\meter} increments and has an initial temperature of \SI{300}{\kelvin}, excited with a Gaussian laser pulse.
The pulse has a full width at half maximum of \SI{50}{\femto\second} and a photon energy of \SI{1.5}{\electronvolt}.
In appendix~\ref{sec:400}, we demonstrate that the signature of the photon energy on the electron distribution is nearly entirely lost after just \SI{25}{\femto\second}. 
Thus, the nonequilibrium state after laser excitation is almost independent of the photon energy.
We excite the sample with various fluences after reflection, ranging from $\Phi = \SI{0.05}{\joule\per\square\meter}$ to $\Phi = \SI{15}{\joule\per\square\meter}$, corresponding to maximum surface electron temperatures from \SI{377}{\kelvin} to \SI{4270}{\kelvin}.

\subsection{Influence of transport on global thermalization}

We examine the evolution of the electron distribution during and after ultrafast laser excitation, focusing on the differences between pure intrinsic scattering and additional transport processes.
To this end, we consider two different spatial excitation profiles, homogeneous and inhomogeneous following Beer-Lambert's law, with the same absorbed fluence.
To obtain a measure of thermalization, we quantify the amount of nonequilibrium by the athermal carrier density (ACD)
\begin{align}
    n_\text{neq}(t, z) = \int D_E \left|f_E(t, z) - f_E^\text{Fermi}(t, z) \right| \dd{E} , \label{eq:nmad}
\end{align}
where the thermal reference distribution $f_E^\text{Fermi}$ is the Fermi-Dirac distribution with the same energy and particle density as the nonequilibrium distribution $f_E$.
In appendix~\ref{sec:matprop}, we describe how we calculate the dynamical electron properties at the points in the material based on the branch-resolved electron distributions.

\begin{figure}
    \includegraphics[trim = 0cm 0cm 0cm 0.865cm, clip]{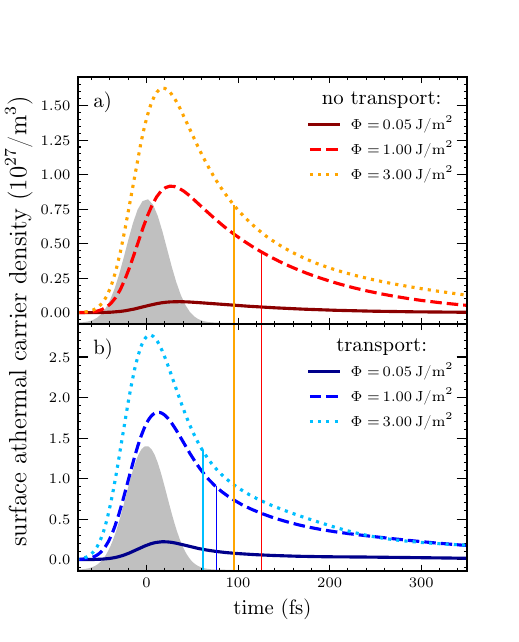}
    \caption{Athermal carrier density (ACD), see \cref{eq:nmad}, at the front surface of the material for different fluences and a simulation a) without transport processes, corresponding to a spatially homogeneous excitation and b) with transport effects, describing an inhomogeneous excitation.
    The temporal laser profile is sketched in gray.
    The time when the density of athermal electrons decays to half of its maximum value is marked with vertical lines.
    For visual clarity, this is omitted for the lowest fluence.
    }\label{fig:dev}
\end{figure}
\Cref{fig:dev} shows the temporal evolution of the ACD directly at the laser-excited surface, within \SI{1}{\nano\meter} depth, for three different excitation fluences and the two different spatial excitation conditions.
In part a), the sample is homogeneously excited, resulting in an electron distribution that is independent of the depth in the material.
Consequently, no transport effects occur in the material.
In part b), the sample is excited to the same energy density, but inhomogeneously according to Beer-Lambert's law, allowing the observation of transport effects.

During laser irradiation, the distribution is driven out of equilibrium, exciting athermal electrons.
Higher fluences lead to a higher density of athermal electrons.
On the following tens to hundreds of femtoseconds, those electrons thermalize.
It is well-known from previous studies that generally higher fluences lead to faster thermalization~\cite{Fann1992a, Mueller2013PRB, Seibel2026arxiv}.
We confirm this and indicate it in part~a) of \cref{fig:dev} for the fluences of \SI{1}{\joule\per\square\meter} and \SI{3}{\joule\per\square\meter} by vertical lines at the time when the ACD has decayed to half of its maximum value.
For part b), the overall dynamics are very similar to part a), but the ACD decays faster for all considered fluences compared to homogeneous excitation because the transport of electrons into the interior of the material adds an additional relaxation channel next to intrinsic thermalization processes.
Thus, the transport process accelerates the effective thermalization at the surface. 

\begin{figure}
    \includegraphics{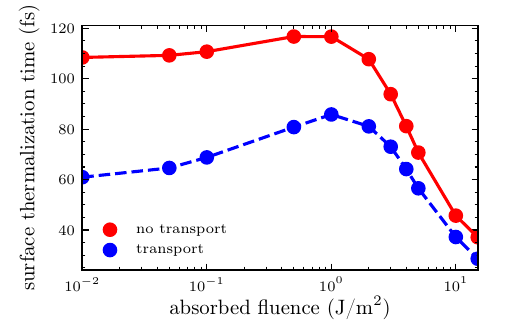}
    \caption{Surface thermalization times extracted from calculations with both spatially homogeneous and inhomogeneous excitation in dependence on absorbed fluence.
    The lines connecting the dots in the figure are provided as a guide to the eye.
    }\label{fig:tau}
\end{figure}
We analyze the influence of electron transport on the timescale of thermalization for a whole range of fluences by extracting an effective surface thermalization time as explained in appendix~\ref{sec:fits}.
\Cref{fig:tau} shows the surface thermalization times in dependence of the absorbed fluence.
For homogeneous excitation, the thermalization time is almost constant at low fluences but decreases drastically for high fluences.
Ref.~\citenum{Seibel2026arxiv} has shown that the rate of thermalization due to electron-electron scattering increases with fluence, while the rate of thermalization due to electron-phonon scattering decreases.
Thus, electron-electron scattering dominates at high fluences.
For low fluences, the two processes cooperate, while in the intermediate fluence regime, electron-phonon scattering hinders electron thermalization, resulting in slightly longer times.

In comparison, inhomogeneous excitation leads to an acceleration of surface thermalization for all fluences.
In particular, when the excitation is weak and electron-electron scattering is therefore slow, the contribution from transport is significant.

\begin{figure}
    \includegraphics[trim = 0cm 0cm 0cm 0.865cm, clip]{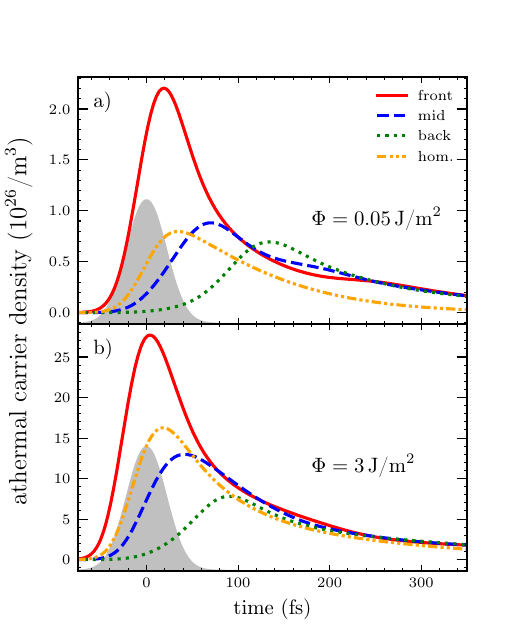}
    \caption{
    Density of athermal carriers at the front surface, back surface, and in the center of the material, calculated with \cref{eq:nmad}, for two different absorbed fluences.
    The result for a homogeneously excited material is further shown for reference.
    The temporal laser profile is sketched in gray.
    }\label{fig:dev_depth}
\end{figure}
However, the motion of athermal electrons away from the surface into the bulk of the material must also affect thermalization dynamics in the bulk, which we will investigate next. 
Therefore, we calculate the evolution of the ACD and show it in \cref{fig:dev_depth} at the front and back surface, as well as in the center of the material for the weakest and strongest excitation investigated in \cref{fig:dev}.
During the laser pulse, the ACD increases at all depths, with decreasing magnitude over depth in the material according to the absorption profile.
At the back of the material, at a depth of \SI{45.5}{\nano\meter}, nearly no athermal electrons are excited by the laser as the remaining intensity is only 0.3\,\% of the front surface value.

After the pulse, when the ACD decreases at the front of the material, it increases significantly at the back due to transport of athermal electrons.
For weak excitation, as shown in panel a), the ACD at the back of the sample eventually surpasses the surface density for a duration of approximately \SI{100}{\femto\second}.
The ACD equilibrates over depth before vanishing, indicating that the equilibration of the spatial gradient is faster than thermalization.

For stronger excitation, shown in \cref{fig:dev_depth}~b), the ACD decreases with depth independently of time, which is a consequence of the faster electron-electron scattering.
Additionally, the transport-mediated equilibration of the spatial gradient is faster for the larger fluence. 

For comparison, \cref{fig:dev_depth} also shows the ACD for homogeneous excitation with the same absorbed fluence.
Here, the maximum athermal carrier density is smaller than for inhomogeneous excitation in the front surface region and larger compared to that at the rear surface of the sample.
Interestingly, the ACD for the homogeneously excited sample drops below the ACD for inhomogeneous excitation at every point in the material already after \SI{100}{\femto\second}.
This seems to contradict the finding of accelerated thermalization presented in \cref{fig:tau}.
However, the thermalization times were determined using the initial decay rate at the surface.
Examining the ACD over the entire depth of the material reveals that electron transport accelerates the decay of athermal carriers at the surface initially, but the presence of the gradient leads to slower thermalization at later times for all depths.
This effect is much more pronounced at low fluence than at high fluence, which is consistent with the varying impact of transport processes, shown in \cref{fig:tau}.

\subsection{Spectral thermalization dynamics}

\begin{figure}
    \includegraphics[trim = 0cm 0cm 0cm 0.865cm, clip]{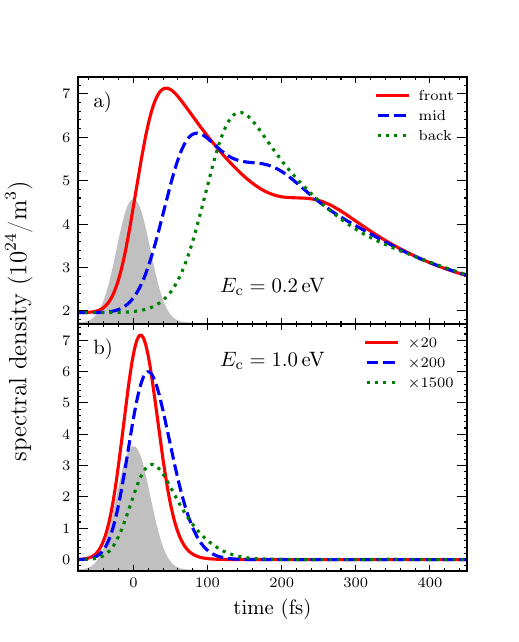}
    \caption{
    Spectral densities, calculated with \cref{eq:spec_dens} at a) $E\ii{c}=\SI{0.2}{\electronvolt}$ and b) $E\ii{c}=\SI{1.0}{\electronvolt}$ above the Fermi level in an energy range with width $\Delta E = \SI{0.1}{\electronvolt}$ at the front surface, back surface and in the center of the material.
    In part b) we scale the spectral densities by factors of $20$ to $1500$ for a better visualization.
    The temporal laser profile is sketched in gray.
    }\label{fig:spec_dens}
\end{figure}
\begin{figure*}
    \centering
    \includegraphics[trim = 0cm 0cm 0cm 0.6cm, clip]{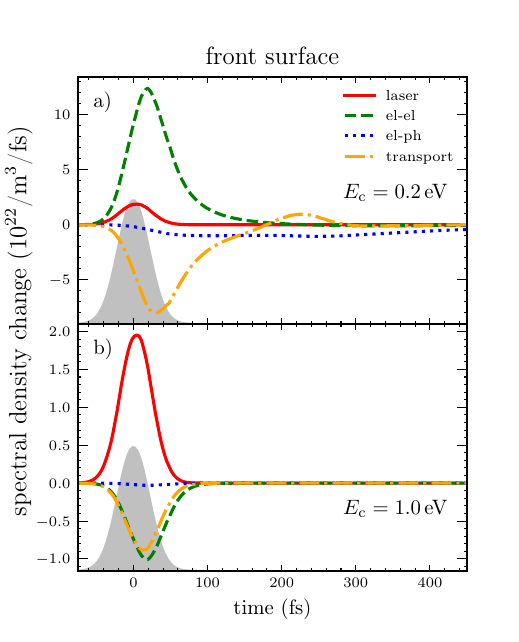}
    \includegraphics[trim = 0cm 0cm 0cm 0.6cm, clip]{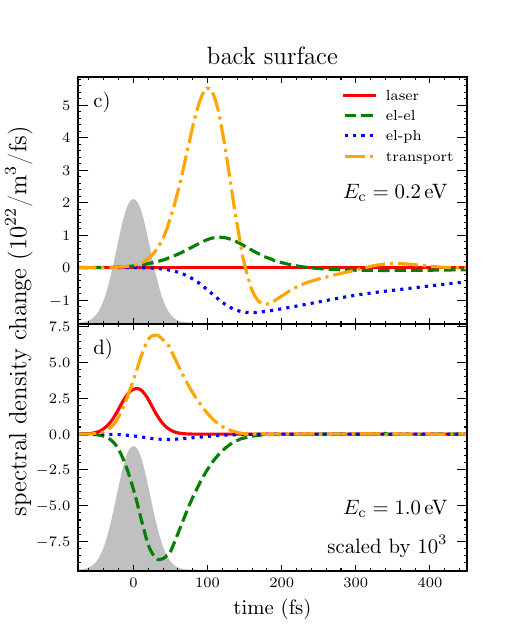}
    \caption{
    Change of the spectral densities, see \cref{eq:spec_dens_change}, at the a) \& b) front surface and c) \& d) back surface due to the various interaction mechanism at \mbox{$E\ii{c}=\SI{0.2}{\electronvolt}$} and \mbox{$E\ii{c}=\SI{1.0}{\electronvolt}$} above the Fermi level in an energy range with width \mbox{$\Delta E = \SI{0.1}{\electronvolt}$}.
    The values for the high central energy at the back surface are scaled by a factor of $10^3$.
    The temporal profile of the laser is sketched in gray.
    }\label{fig:spec_dens_change}
\end{figure*}

So far we have analyzed the ACD, which traces the total amount of athermal electrons.
However, it is an integrated quantity that cannot capture the details of the intrinsically energy-dependent dynamics of the nonequilibrium distribution~\cite{Bauer2015,Seibel2023}. 
Experimentally, this energy-dependent behavior of thermalization is accessible through photoemission spectroscopy~\cite{Bauer2015,Kuehne2022,Beyazit2020}, where measured intensities are related to the density of electrons in a small energy range~\cite{Uehlein2022}. 
We determine these so-called spectral densities
\begin{align}
    n_{E_\text{c}, \Delta E}(t, z) &= \int_{E\ii{c}-\Delta E/2}^{E\ii{c}+\Delta E/2} f_E(t, z)\,D(E) \dd{E}\,, \label{eq:spec_dens}
\end{align}
in an energy interval with a width of $\Delta E$ around a central energy $E\ii{c}$. 
\Cref{fig:spec_dens} shows the spectral densities at two central energies and three depths in the material for the fluence of $\Phi = \SI{0.05}{\joule\per\square\meter}$, which is in an excitation range where we expect a large influence from transport effects, see \cref{fig:tau,fig:dev_depth}.

\Cref{fig:spec_dens}~a) depicts the spectral densities at \mbox{$E\ii{c}=\SI{0.2}{\electronvolt}$}, directly above the Fermi edge.
During the laser pulse, the spectral densities at the front surface increase due to excitation of electrons by the laser and secondary electron generation.
In the center and in the back of the material, the spectral densities increase with a time delay, indicating the influence of transport processes on the spectral densities.
This aligns with the results of photoemission experiments that compare the intensities of front and back probes~\cite{Kuehne2022}.
We observe a higher maximum of spectral densities at the back than in the center of the material.
This is due to a transiently accumulation of electrons, when they are reflected at the back surface.
This is described in more detail in \cref{sec:neq} of the appendix.

In the following hundreds of femtoseconds, the spectral densities decrease due to scattering and transport processes, which is discussed below.
After \SI{350}{\femto\second}, the spectral densities are homogeneous throughout the material.

The spectral densities at the higher energy, \mbox{$E\ii{c} = \SI{1.0}{\electronvolt}$}, shown in part b) are much smaller than at the lower energy and drastically decrease with increasing depth.
At all depths, the spectral densities rise on the timescale of irradiation and decay rapidly with only a slight time delay between their maximum value at increasing depth.

We unravel the origin of the distinct behavior of the spectral densities by separating the influence of the various interaction mechanisms.
Analogously to \cref{eq:spec_dens}, we can calculate the change in spectral densities due to different interaction mechanisms
\begin{align}
    \left.\pdv{n(t, z)}{t}\right|_{\stackrel{{E_\text{c}, \Delta E,}}{\text{inter}}} &= \int_{E\ii{c}-\Delta E/2}^{E\ii{c}+\Delta E/2} \left.\pdv{f_E(t, z)}{t}\right|\ii{inter}\,D(E) \dd{E}\,, \label{eq:spec_dens_change}
\end{align}
by integrating the temporal change of the energy distribution obtained from the individual collision integrals and transport terms in \cref{eq:boltzmann}.

In \cref{fig:spec_dens_change}, we evaluate the changes in spectral density due to laser excitation, electron-electron scattering, electron-phonon scattering, and transport at the front and back surfaces for the two central energies considered in \cref{fig:spec_dens}.
For the energy close to the Fermi edge in panel a), the dynamics at the front surface are mainly driven by electron-electron scattering and transport processes away from the surface, both contributing with opposite signs.
After \SI{200}{\femto\second}, the transport contribution changes sign because electrons are transported back towards the front surface, resulting in the plateau during the decrease of the spectral densities seen in \cref{fig:spec_dens}~a).
\Cref{fig:spec_dens_change}~b) shows the change in spectral density at the high central energy of \SI{1}{\electronvolt}. 
Here, the spectral density increases due to laser absorption and decreases almost equally due to electron-electron scattering and transport processes. 
Note that the change due to laser absorption is nearly identical in parts a) and b), however, the other contributions differ strongly in magnitude.
At the front surface, the dynamics at the high central energy occur almost entirely on the timescale of the laser pulse due to the fast electron-electron scattering.

\Cref{fig:spec_dens_change}~c) and d) show the spectral density change at the back surface for the two central energies.
Laser excitation is almost negligible due to the weak intensity of the laser at the back surface.
Instead, the spectral densities at the Fermi edge are mainly driven by transport processes.

At the high central energy, the increase in spectral density due to transport is almost as small as direct laser excitation, in particular, three orders of magnitude smaller than at the low central energy in part c).
The increase is counteracted by efficient recombination due to scattering.

In all cases, electron-phonon scattering decreases the spectral density. However, its influence is noticeable only at low energies at the back surface.

In summary, we find that mainly electron transport drives the dynamics near the Fermi edge, with the majority of electrons reaching that energy region through secondary electron generation induced by scattering.

\section{Conclusion}

We have examined the influence of transport on thermalization in laser-driven metals in dependence on time, space, and energy.
To that end, we have reformulated the Boltzmann equation in energy space to include both spatial transport and intrinsic scattering with full collision integrals.
We have compared the dynamics of the electron distribution in an aluminum-like metal emerging from spatially homogeneous and inhomogeneous excitation at the same absorbed fluence.

We have found that transport accelerates the effective thermalization at the irradiated surface with the strongest effect at low fluences.
Nevertheless, in the later stages of thermalization, the presence of spatial gradients leads to a larger persistent athermal carrier population across the entire sample.
We have found that the electron system does not fully thermalize until the spatial gradient is fully equilibrated, while the spatial equilibration due to transport effects accelerates with excitation strength. 

Connecting to experimentally accessible signals, we have further evaluated the thermalization dynamics in different narrow energetic regions above the Fermi edge.
We have disentangled the contributions of transport and various scattering processes to the dynamics of these spectral densities.
Here, we have found that, depending on both, depth in the material and the energetic position, the spectral density dynamics can be governed either by local electron-electron scattering or non-local transport processes.

Overall, our results show that electronic thermalization in optically thick laser-excited metals cannot be interpreted as a purely local process.
Surface-sensitive observables may indicate rapid thermalization because transport removes athermal carriers from the probed region, while the full electron system remains in nonequilibrium due to spatial redistribution.
This distinction is particularly important for energy-resolved ultrafast measurements, where the dominant relaxation channel depends on both the probed energy window and the probing depth.
The presented framework therefore provides a microscopic route to connect transport, scattering, and experimentally accessible hot-carrier dynamics in laser-excited metals relevant for hot-electron-based applications.

\section*{Acknowledgements}
We acknowledge support through the Deutsche Forschungsgemeinschaft (DFG, German Research Foundation) - TRR 173 - 268565370 Spin+X (project no. A08 and INF).
We appreciate the Allianz für Hochleistungsrechnen Rheinland-Pfalz for providing computing resources through project STREMON on the Elwetritsch high-performance computing cluster.

\section*{Data availability}
The data that support the findings of this study are openly available at the following URL/DOI~\cite{Uehlein2026zenodo}:\\\url{https://doi.org/10.5281/zenodo.20512643}.

\clearpage
\bibliographystyle{IEEEtran}            
\bibliography{bibfile/all.bib}

\clearpage
\appendix
\section{Averaged velocity}\label{sec:vel}
The velocities of electrons with velocity $v$ moving towards the back side of the material are equally distributed over the surface of a hemisphere.
In spherical coordinates the $z$-component is given by $v_z = v \cos(\vartheta)$.
The average over the hemisphere results in
\begin{align}
    \bar{v}_z &= \frac{1}{A} \int_{0}^{2\pi} \int_{0}^{\pi/2} v_z\,v^2 \sin(\vartheta) \, \dd\vartheta \, \dd\varphi = \frac{v}{2}
\end{align}
with the surface of a half sphere $A=2\pi v^2$.
The same applies to the velocity of electrons moving in the opposite direction.

\section{Influence of electron-phonon scattering on temperature equilibration}\label{sec:neq}
\begin{figure}
    \includegraphics{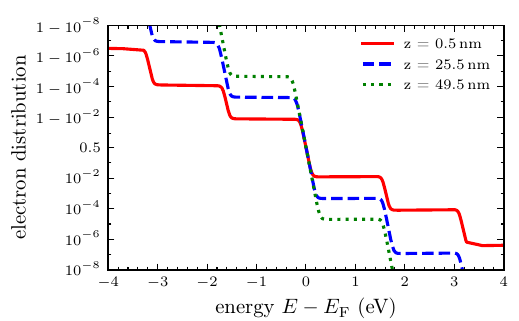}
    \caption{Initial nonequilibrium distributions in various depth in the material in a symmetric logarithmic representation. 
    }\label{fig:dist_exec}
\end{figure}
To investigate how various scattering processes affect the decay of an initial energy gradient, we will decouple the process of laser excitation and thermalization.
Therefore, we set up our calculations with nonequilibrium distributions shown in \cref{fig:dist_exec}.
A typical step-like structure, with a step width corresponding to the photon energy of $\hbar\omega = \SI{1.5}{\electronvolt}$, can be observed.
We apply a fluence after reflection of $\Phi = \SI{1}{\joule\per\meter\squared}$.
The height of the steps correlates with the intensity.
For this reason, the height of the steps decreases exponentially with the distance from the surface.
Please note, an exponential decrease appears linear on a logarithmic scale.

\begin{figure}
    \includegraphics{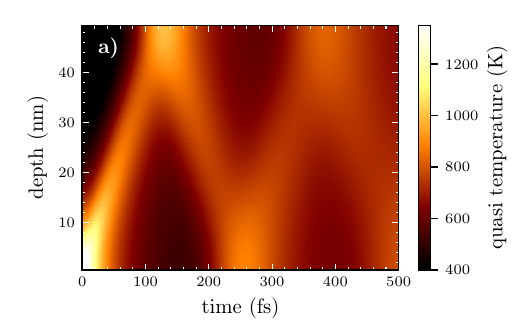}
    \includegraphics{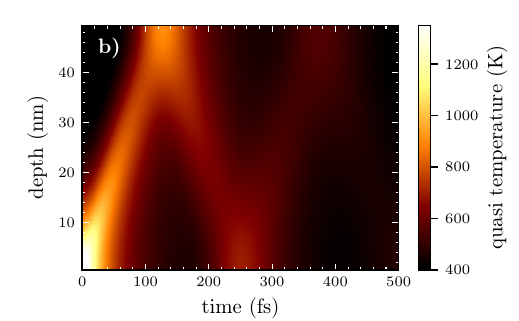}
    \includegraphics{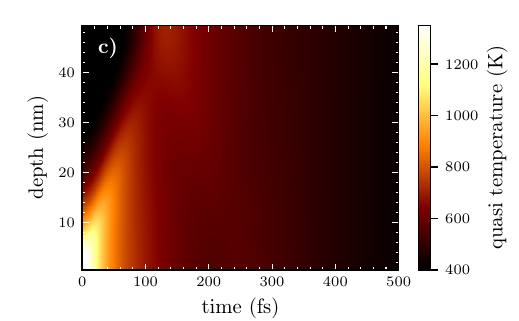}
    \caption{Influence of different assumptions for electron-phonon scattering on the propagation of energy through depth and time shown at the quasi temperature of the electrons: a) no electron-phonon collisions, b) only electron-phonon collisions without change of direction in the electrons, c) all electron-phonon collisions. 
    The calculations are initialized with an exponential energy gradient in the nonequilibrium distribution as shown in \cref{fig:dist_exec}.
    Temperatures below \SI{400}{\kelvin} and above \SI{1350}{\kelvin} are not resolved in the colorbar.
    }\label{fig:T_maps}
\end{figure}
The electrons do not follow a Fermi distribution, for this reason, no temperature is defined on the timescales of electron thermalization. 
Nevertheless, it can be useful to define a quasi temperature for electrons in nonequilibrium as well, to obtain a measure for the energy content of the system and to illustrate the mutual influence of transport and local scattering. 
We define the quasi temperature $T(t, z)$ as the temperature of a Fermi-Dirac distribution that has the same energy and particle density as the nonequilibrium distribution. Further details can be found in appendix~\ref{sec:matprop}.

In \cref{fig:T_maps}, we evaluate the quasi temperature of the electrons in dependence of time and depth in the material.
To examine the influence of electron-phonon scattering on the equilibration of the initial energy gradient, we consider three cases:
no electron-phonon collisions at all, only electron-phonon collisions without change of direction in the electrons, and all electron-phonon collisions.
In \cref{fig:T_maps}~a) we show the quasi temperature for a calculation where electron-phonon scattering is ignored.
The observed exponential decay in temperature dissipates within a few femtoseconds.
A high-temperature pulse passes through the material and reflects off its surfaces.
Due to the energy exchange during electron-electron scattering between the two branches, the pulse diverges slowly.
Such oscillations were observed in reflectivity experiments for gold~\cite{Liu2005}.
Here, transport effects are overestimated by neglecting electron-phonon scattering.

In \cref{fig:T_maps}~b) and c) we show results of calculations that include electron-phonon scattering, where in part~b) only electron-phonon scattering, in which the electrons retain their direction of motion, is included, while in part~c) all types of electron-phonon scattering are captured.
Both parts exhibit similar oscillations to part~a), but with greater damping.
The influence of the direction change of the electrons in part~c) results in a even greater divergence of the pulse.

Thus, the ratio of electron-phonon collisions that cause the electrons to change direction determines how strongly the oscillation is damped by the material.
We set the proportion of collisions that cause a change in direction to a rather large value of 30\% to counteract the omitted velocity dependence of the distribution, which overestimates the effect of transport.
While the magnitude of the damping depends on this parameter, the qualitative behavior remains unaffected.

Since the simultaneous dependence on space and time often makes the presentation of data in the form of graphs difficult to understand, we have published a series of videos that show the temporal development of the quasi temperature plotted over space~\cite{Uehlein2026zenodo}.
We show, for example, videos of the calculations presented in \cref{fig:T_maps} and additional calculations for a \SI{150}{\nano\meter} thick material.

\section{Comparison to two-temperature model}\label{sec:ttm}
\begin{figure}[t]
    \includegraphics{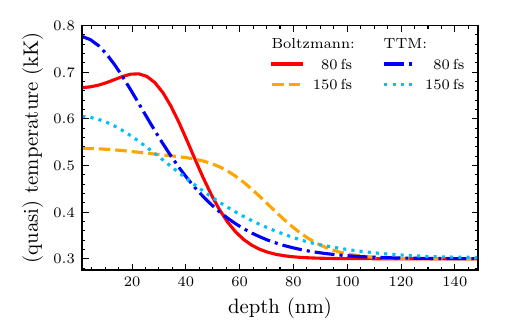}
    \caption{Temperature in dependence of depth at \SI{80}{\femto\second} and \SI{150}{\femto\second}.
    We compare results obtained from a Boltzmann calculation with results obtained from a TTM calculation.
    }\label{fig:ttm}
\end{figure}

We compare the results obtained from a Boltzmann calculation with the results obtained from a two-temperature model (TTM) calculation.
We implement the two-temperature model as described in Refs.~\citenum{Rethfeld2017} and \citenum{Uehlein2025}.
We assume a constant thermal conductivity of \mbox{$\kappa = \SI{237}{\joule\per\meter\per\second\per\kelvin}$~\cite{Kittel}}.
The calculations are initialized with an exponential temperature profile matching the energy density of the distributions in \cref{fig:dist_exec}.

\Cref{fig:ttm} shows the temperature in dependence of depth at \SI{80}{\femto\second} and \SI{150}{\femto\second}.
We published a video showing the whole time evolution of the temperature in dependence of depth~\cite{Uehlein2026zenodo}.
In the calculation with the Boltzmann model, it is visible that a heat wave travels through the material, while in the calculation with the TTM, the temperature gradient decreases steadily.
This shows the infinite speed of heat propagation in the TTM in contrast to the finite speed of heat transport in the Boltzmann model.
Taking a finite velocity into account by extending the TTM using Cattaneo's law can also produce a heat wave similar to our results in a temperature-based model~\cite{Klossika1996}. 
The time of equilibration of an initial temperature gradient is comparable in both models.

\section{Density of states}\label{sec:dos}
\begin{figure}
    \includegraphics{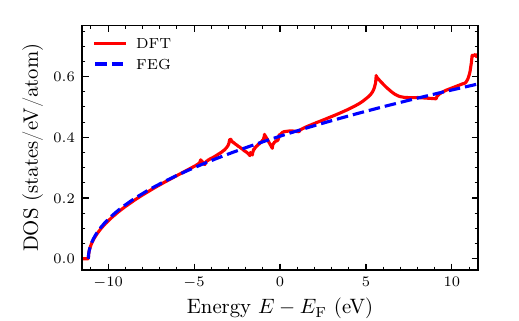}
    \caption{DOS of a FEG in comparison to the DOS of aluminum calculated with density functional theory~\cite{Uehlein2025}. The effective mass was chosen to match the particle density of aluminum.}\label{fig:dos}
\end{figure}

The DOS in the free electron gas is given by
\begin{align}
    D(E) = \frac{\sqrt{2 m^3}V}{\pi^2 \hbar^3} \sqrt{E}\,,
\end{align}
with the effective mass $m$.
\Cref{fig:dos} shows the DOS of aluminum calculated from density functional theory~\cite{Uehlein2025} and the FEG, which was used in this manuscript.
We choose $m = \SI{1.05}{m_e}$, where $m_e$ is the electron mass, to match the particle density with the Fermi energy extracted from the DFT calculation.

\section{Influence of exciting photon energy}\label{sec:400}
\begin{figure}
    \includegraphics{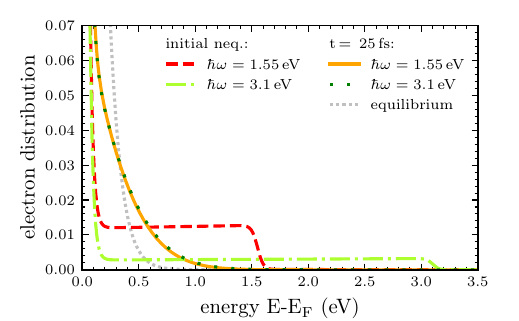}
    \caption{Electron distribution at \SI{25}{\femto\second} compared to an initial nonequilibrium for calculations with different photon energies~$\hbar\omega$. 
    At \SI{25}{\femto\second} the corresponding equilibrium distribution is shown.
    }\label{fig:dist}
\end{figure}
\Cref{fig:dist} shows an initial nonequilibrium distribution of the electrons on the material's surface, as well as the distributions \SI{25}{\femto\second} later.
Additionally, the corresponding equilibrium distribution at \SI{25}{\femto\second} is shown in gray.
We compare calculations for the photon energy of $\hbar\omega = \SI{1.55}{\electronvolt}$ with calculations for $\hbar\omega = \SI{3.1}{\electronvolt}$.
Except for the photon energy, all other parameters are the same.
The initial steps with the width of the photon energy are clearly visible.
Since both distributions contain the same energy, the height of the step for the \SI{3.1}{\electronvolt} excitation is lower than for the \SI{1.55}{\electronvolt} excitation.
Within \SI{25}{\femto\second}, both distributions thermalize to a nearly identical distribution.
This distribution does not exhibit any characteristics of the step, but differs significantly from the corresponding equilibrium distribution.
We published a video showing the time evolution of this distributions~\cite{Uehlein2026zenodo}.

\begin{figure}[t]
    \includegraphics[trim = 0cm 0cm 0cm 0.865cm, clip]{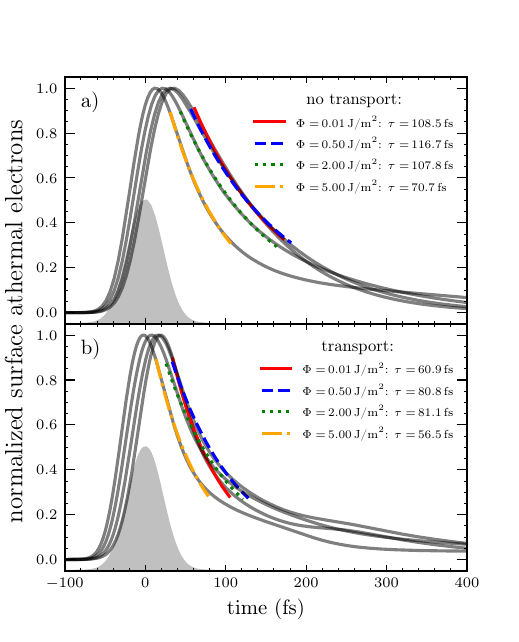}
    \caption{Deviation from equilibrium shown as the normalized ACD at the front surface for different fluences $\Phi$ for a) homogeneous excitation and b) inhomogeneous excitation.
    An exponential fit is applied to the decay of the deviation to define a thermalization time $\tau$.
    The temporal laser profile is sketched in gray.
    }\label{fig:taus}
\end{figure}

\section{Calculation of electron properties}\label{sec:matprop}
We calculate the time evolution of the electron distributions $f^\pm$ for the two dispersion branches at different depths in the material.
We calculate the energy-, space- and time-resolved population density
\begin{align}
    D_E f_E(t,z) = \sum_{\pm} D^\pm_E f^\pm_E(t,z)
\end{align}
by summing over the contributions of both branches.
This also defines the particle and energy densities and therefore enables the calculation of the corresponding thermal distribution, which also defines a quasi temperature~$T(t, z)$ and quasi chemical potential~$\mu(t, z)$, see Ref.~\citenum{Uehlein2025} for further details.

\section{Extraction of thermalization time}\label{sec:fits}
We define the deviation from the equilibrium as the normalized ACD, see \cref{eq:nmad}, and investigate the influence of the fluence on the thermalization time by extracting the thermalization time from an exponential fit to the decay of the deviation at the front surface.
\Cref{fig:taus} shows the deviation from the equilibrium and the fits to it for a) homogeneous excitation, thus no transport effects occur, and b) inhomogeneous excitation, where transport processes influence the dynamics.
We use the part of the decay where the deviation is between 0.9 and 0.3 to reduce the influence of the laser pulse and the phonons, as well as of transport processes back to the excited surface.
It's evident that the thermalization is faster for inhomogeneous excitation than for homogeneous excitation.
\vspace{2.5cm}

\end{document}